\documentclass[aps,twocolumn,showpacs,preprintnumbers,noshowpacs,superscriptaddress,6pt]{revtex4-2}

\usepackage[utf8]{inputenc}
\usepackage{graphicx}
\usepackage{subfigure}
\usepackage{multirow}
\usepackage{adjustbox}
\usepackage{natbib}

\usepackage{longtable}
\usepackage{parskip}
\usepackage[T1]{fontenc}
\usepackage{dcolumn}

\usepackage{bm}
\usepackage{amsfonts}
\usepackage{amsmath}
\usepackage{amssymb}
\usepackage{siunitx}
\usepackage{booktabs}
\usepackage{physics}
\usepackage{mathtools}
\usepackage{flafter}
\usepackage{threeparttable}
\usepackage{subfiles}
\usepackage{subfigure}
\usepackage{import}
\usepackage{float}
\usepackage{makecell}
\usepackage{diagbox}

\usepackage{ulem}

\usepackage{colortbl}
\usepackage{indentfirst}

\begin{document}

\title{Ionization potentials and electron affinities of Rg, Cn, Nh, and Fl superheavy elements}

\author{M.~Y.~Kaygorodov}
\affiliation{Department of Physics, St. Petersburg State University, 7/9 Universitetskaya nab., 199034 St. Petersburg, Russia}

\author{D.~P.~Usov}
\affiliation{Department of Physics, St. Petersburg State University, 7/9 Universitetskaya nab., 199034 St. Petersburg, Russia}

\author{E.~Eliav}
\affiliation{School of Chemistry, Tel Aviv University, 69978 Tel Aviv, Israel}

\author{Y.~S.~Kozhedub}
\affiliation{Department of Physics, St. Petersburg State University, 7/9 Universitetskaya nab., 199034 St. Petersburg, Russia}

\author{A.~V.~Malyshev}
\affiliation{Department of Physics, St. Petersburg State University, 7/9 Universitetskaya nab., 199034 St. Petersburg, Russia}

\author{A.~V.~Oleynichenko}
\affiliation{B. P. Konstantinov Petersburg Nuclear Physics Institute of National Research Centre ``Kurchatov Institute'', Gatchina, 188300 Leningrad District, Russia}

\author{V.~M.~Shabaev}
\affiliation{Department of Physics, St. Petersburg State University, 7/9 Universitetskaya nab., 199034 St. Petersburg, Russia}

\author{L.~V.~Skripnikov}
\affiliation{B. P. Konstantinov Petersburg Nuclear Physics Institute of National Research Centre ``Kurchatov Institute'', Gatchina, 188300 Leningrad District, Russia}
\affiliation{Department of Physics, St. Petersburg State University, 7/9 Universitetskaya nab., 199034 St. Petersburg, Russia}

\author{A.~V.~Titov}
\affiliation{B. P. Konstantinov Petersburg Nuclear Physics Institute of National Research Centre ``Kurchatov Institute'',
Gatchina, 188300 Leningrad District, Russia}

\author{I.~I.~Tupitsyn}
\affiliation{Department of Physics, St. Petersburg State University, 7/9 Universitetskaya nab., 199034 St. Petersburg, Russia}

\author{A.~V.~Zaitsevskii}
\affiliation{B. P. Konstantinov Petersburg Nuclear Physics Institute of National Research Centre ``Kurchatov Institute'',
Gatchina, 188300 Leningrad District, Russia}
\affiliation{Department of Chemistry, M. V. Lomonosov Moscow State University, 119991 Moscow, Russia}


\begin{abstract}
The successive ionization potentials (IPs) and electron affinities (EAs) for superheavy elements with~$111 \leq Z \leq 114$, namely, Rg, Cn, Nh, and Fl  are reexamined using the relativistic Fock-space coupled-cluster method with nonperturbative single (S), double (D), and triple (T) cluster amplitudes (FS-CCSDT).
For the most of considered quantities, the triple-amplitude contributions turn out to be important.
The Breit and frequency-dependent Breit corrections are evaluated by means of the configuration-interaction method.
The quantum-electrodynamics corrections to the IPs and EAs are taken into account within the model-QED-operator approach.
The obtained results are within~$0.10$~eV uncertainty. 
\end{abstract}

\maketitle

\section{Introduction}\label{sec:intro}
Superheavy elements (SHEs) are trans-uranium elements which are artificially synthesized in the cold- and hot-fusion nuclear reactions with neutron-rich isotopes, see, e.g., reviews~\cite{2017_OganessianY_PhysScrip, 2019_GiulianiS_RevModPhys, 2019_DullmannC_RadiochimActa}.
All the known SHEs, starting from rutherfordium~$(Z\!\!=\!\!104,\mathrm{Rf})$ and ending with oganesson~$(Z\!=\!118,\mathrm{Og})$, belong to the seventh row of the periodic table; here and below~$Z$ is the nuclear charge number.
The experimental study of SHE electronic-structure properties is exceptionally difficult due to the extremely low production rates and short lifetimes.
Nowadays, the heaviest elements for which experimental data on such electronic-structure property as ionization potential exist are nobelium~$(Z\!=\!102,\mathrm{No})$~\cite{2016_LaatiaouiM_Nature, 2017_ChhetriP_EurPhysJD,2018_ChhetriP_PhysRevLett,2018_SatoT_JAmChemSoc} and lawrencium~$(Z\!=\!103,\mathrm{Lr})$~\cite{2015_SatoT_Nature,2018_SatoT_JAmChemSoc}.
However, promising experimental techniques~\cite{2016_OganessianY_RussChemRev,2016_LautenschlagerF_NIMB,2020_LaatiaouiM_PhysRevLett} may expand our knowledge on the electronic structure of the heavier elements beyond lawrencium.
\par
Modeling of the electronic structure of SHEs is a challenging task for the modern theoretical atomic physics.
The necessity to explicitly account for correlations between a large number of electrons demands huge computational efforts.
Moreover, very strong electric fields induced by heavy nuclei make relativistic and quantum-electrodynamics (QED) effects substantial.
As a result, to predict properties of SHEs with reliable accuracy one has to employ an efficient method for evaluation of the electron-correlation effects within the relativistic framework.
\par
Some SHEs have been studied previously employing various methods, such as the relativistic Fock-space coupled clusters with single (S) and double (D) amplitudes (FS-CCSD)~\cite{1994_EliavE_PhysRevLett,1995_EliavE_PhysRevA,1996_EliavE_PhysRevA,1996_EliavE_PhysRevLett,2001_LandauA_JChemPhys,2015_BorschevskyA_PhysRevA,2019_KahlE_PhysRevA}, the configuration-interaction method combined with the many-body perturbation theory (CI\texttt{+}PT) approach~\cite{2008_DinhT_PhysRevA78, 2016_DzubaV_HyperfInteract237, 2016_DzubaV_PhysRevA,2016_GingesJ_PhysRevA,*2016_GingesJ_JPhysBAtMolOptPhys, 2018_LackenbyB_PhysRevA98_Db, 2018_LackenbyB_PhysRevA98_Og,2020_LackenbyB_PhysRevA101, 2021_AllehabiS_PhysRevA}, the multiconfiguration Dirac-Fock method (MCDF)~\cite{2007_YuY_EurPhysJD, 2008_YuY_JChemPhys}.
In general, all above-mentioned calculations are in agreement with each other, however, in some cases there is a discrepancy between the FS-CCSD and CI\texttt{+}PT results.
For example, the discrepancy is about~$0.2$~eV for the first IP of copernicium~($Z=112$, Cn)~\cite{2001_LandauA_JChemPhys, 2020_LackenbyB_PhysRevA101} and $0.5$~eV for the same quantity of roentgenium~($Z=111$, Rg)~\cite{1994_EliavE_PhysRevLett, 2020_LackenbyB_PhysRevA101}.
These inconsistencies have served as motivation to thoroughly reexamine the electronic-structure properties of the SHEs.
\par
The present paper aims to revise ionization potentials (IPs) and electron affinities (EAs) of SHEs with~$111\leq Z \leq 114$ by means of the relativistic FS-CC method, with the fully-relativistic Dirac-Coulomb Hamiltonian being employed.
The SD and nonperturbative triple (T) cluster amplitudes are considered in the calculations.
In most cases, the triple-amplitude corrections, which are evaluated for the first time for these systems, turn out to be comparable or even larger than the uncertainties of the FS-CCSD correlation treatment.
A basis set used to solve the FS-CCSD equations is optimized for each particular element in order to minimize the basis-set restriction error.
The FS-CC results are supplemented by the Gaunt, retardation, and frequency-dependent Breit corrections as well as the QED contributions. 
These effects are evaluated separately using the configuration-interaction (CI) method in the basis of the Dirac-Fock-Sturm (DFS) orbitals and the model-QED-operator approach. 
We demonstrate the consistency of adding the CI-DFS based corrections to the FS-CC results by using the Gaunt interaction as an example.
\par
The paper has the following structure. In Sec.~\ref{sec:methods}, a brief introduction to the FS-CC and CI-DFS methods is presented and key features of their implementation are overviewed.
In Sec.~\ref{sec:calculations}, the thorough details of the calculations are given.
The obtained results are discussed and compared with values available in the literature in Sec.~\ref{sec:results}, with each element being placed in a separate subsection for convenience.
\par
Atomic units are used throughout the paper.

\section{Methods}\label{sec:methods}
In the present work, the electronic structure of the SHEs is calculated by means of a combination of the FS-CC and CI methods. 
Namely, the interelectronic interaction represented by the Dirac-Coulomb Hamiltonian is evaluated within the FS-CC approach.
Various corrections to the FS-CC values are calculated using the CI method.
\par
Within the FS-CC method~\cite{1991_KaldorU_TheoretChimActa,2001_VisscherL_JChemPhys,2017_EliavE_HandbookofRelativisticQuantumChemistry}, the model space (MS) is defined through a projection operator that is decomposed into the Fock-space sector projectors corresponding to different numbers of particles~$(p)$ or holes~$(h)$ above the Fermi-vacuum state.
The latter one is considered as the Fock-space sector~$0h0p$.
It is most convenient to use the closed-shell Fermi-vacuum state.
The FS-CC equations are formulated for an effective many-electron Hamiltonian~$H_{\mathrm{eff}}$ which is associated with some Hamiltonian~$H$, e.g., the Dirac-Coulomb one, and operates in the MS.
The valence-universal wave operator~$\Omega$ is constructed successively for each sector of the Fock space starting from the sector~$0h0p$.
For the latter, the FS-CC equations are reduced to the single-reference CC ones.
In contrast to the single-reference CC method, the FS-CC method allows one to consider static correlations in terms of the MS-state mixing.
\par
The FS-CC method implemented in the EXP-T program~\cite{2020_OleynichenkoA_Symmetry,Oleynichenko_EXPT,EXPT_website} is used in the present work.
We adopt the notation FS-CCSD for the calculations involving the single (S) and double (D) cluster amplitudes and FS-CCSDT for the calculations involving SD and nonperturbative triple (T) amplitudes.
The Dirac-Fock calculations, otherwise known as the relativistic Hartree-Fock ones, and subsequent integral-transformation steps are carried out using the DIRAC package~\cite{DIRAC19, 2020_SaueT_JChemPhys}.
\par
\par
The relativistic Dirac-Coulomb many-electron Hamiltonian $H_{\mathrm{DC}}$ is formulated for the present FS-CC calculations as follows:
\begin{equation}\label{eq:H^DC}
    H_{\mathrm{DC}} = \Lambda^{+} \left[ H^{\mathrm{D}} + V^{\mathrm{C}} \right] \Lambda^{+},
\end{equation}
where $H^{\mathrm{D}}$ is the sum of the one-electron Dirac Hamiltonians, $V^{\mathrm{C}}$ is the two-electron Coulomb-interaction operator, and $\Lambda^{+}$ is the projector on the positive-energy states of the Dirac-Fock (DF) Hamiltonian $h^{\mathrm{DF}}$.
The Gaussian model for the nuclear-charge density is employed to describe the nuclear potential.
\par
Within the current implementation of the FS-CC method, the Gaunt-interaction operator~$V^{\mathrm{G}}$ can be included into the calculations by means of the exact two-component molecular mean-field Hamiltonian~$H_{\mathrm{X2Cmmf}}$ approach~\cite{2009_SikkemaJ_JChemPhys}.
This approach allows one to construct the~two-component Hamiltonian, which exactly reproduces the relativistic DF positive-energy eigenvalues.
In the transformation of the Hamiltonian to its second-quantized form required for the correlated calculations, the Gaunt-interaction operator~$V^{\mathrm{G}}$ can be incorporated through a correction to the mean field.
\par
The CI calculations are based on the CI-DFS method~\cite{2003_TupitsynI_OptSpectrosc, 2003_TupitsynI_PhysRevA, 2005_TupitsynI_PhysRevA, 2018_TupitsynI_PhysRevA}.
In this method, the active space (AS) of one-electron orbitals consists of numerical solutions of the DF and DFS equations considered in the central-field approximation.
The CI-correlated many-electron wave function~$\Psi(JM)$ with the total angular momentum~$J$ and its projection~$M$ is represented as a sum of configuration-state functions.
The latter are eigenfunctions of the operator~$J^2$ and are constructed from the orbitals included into the AS according to the restricted active-space (RAS) scheme~\cite{1988_OlsenJ_JChemPhysB}.
\par
The Dirac-Coulomb-Breit Hamiltonian is used in the CI-DFS method:
\begin{equation}\label{eq:H^DCB}
    H_{\mathrm{DCB}} = {\Lambda}^{+} \left[ H^{\mathrm{D}} + V^{\mathrm{C}} + V^{\mathrm{G}} + V^{\mathrm{R}} \right] {\Lambda}^{+},
\end{equation}
where~$V^{\mathrm{R}}$ is the retardation part of the Breit-interaction operator.
The energies corresponding to the states~$\Psi(JM)$ are given by the eigenvalues of the Hamiltonian represented in the many-electron basis of the configuration-state functions.
\par
The QED effects related to the vacuum-polarization (VP) and self-energy (SE) corrections can be incorporated into the CI-DFS calculations by means of the model-QED-operator approach presented in Refs.~\cite{2013_ShabaevV_PhysRevA, 2015_ShabaevV_CompPhysComm, *2018_ShabaevV_CompPhysComm}.
The approach consists in the inclusion of the model-QED operator~$V^{\mathrm{QED}}$ into the one-electron part of the Hamiltonian $H_{\mathrm{DCB}}$ together with the inclusion of the same operator into the DF and DFS equations.
The SE part of the operator~$V^{\mathrm{QED}}$ is constructed from the corresponding matrix elements precalculated for hydrogen-like ions, whereas the VP part is given by the local one-electron potentials. 
Therefore,~$V^{\mathrm{QED}}$ exactly reproduces the lowest-order QED corrections in a system without the interelectronic interaction.
The higher-order QED effects are accounted for only approximately. 
The frequency-dependent Breit-interaction correction is evaluated by using the full Coulomb-gauge interelectronic-interaction operator (see, e.g., Ref.~\cite{2013_ShabaevV_PhysRevA} and references therein) instead of the Coulomb $V^{\mathrm{C}}$ and standard Breit $V^{\mathrm{G}} + V^{\mathrm{R}}$ interactions.  
\par

\section{Computational details}\label{sec:calculations}
The dominant contribution to the IPs and EAs of the elements with~$111\leq Z \leq 114$ is computed with the FS-CC method.
The closed-shell configuration~$[\mathrm{Rn}]5f^{14}6d^{10}7s^2$, which appears to be the ground-state configuration of neutral Cn, is chosen as the sector~$0h0p$.
The detailed FS-CC sector-partitioning schemes used in the calculations are the following:
\\
\begin{eqnarray}
\mathrm{Rg}^{+}(2h0p) \xleftarrow{} \mathrm{Rg}^{}(1h0p)  \xleftarrow[]{} \mathrm{Rg}^{-}(0h0p),\label{eq:z111_FS-CC_sceme} \\[10pt]
\mathrm{Cn}^{2+}(2h0p) \xleftarrow{} \mathrm{Cn}^{+}(1h0p)  \xleftarrow[]{} \mathrm{Cn}^{}(0h0p),\label{eq:z112_FS-CC_sceme} \\[10pt]
\begin{aligned}
    \mathrm{Nh}^{3+}(2h0p) \xleftarrow{}  \mathrm{Nh}^{2+}(1h0p)  \xleftarrow[]{} \mathrm{Nh}^{+}(0h0p) \xrightarrow[]{} \\ 
    \xrightarrow[]{}   \mathrm{Nh}^{}(0h1p)  \xrightarrow[]{} \mathrm{Nh}^{-}(0h2p)\label{eq:z113_FS-CC_sceme},
\end{aligned}
\\[10pt]
\begin{aligned}
    \mathrm{Fl}^{4+}(2h0p) \xleftarrow{} \mathrm{Fl}^{3+}(1h0p)  \xleftarrow[]{} \mathrm{Fl}^{2+}(0h0p)  \xrightarrow[]{} \\ \xrightarrow[]{}   \mathrm{Fl}^{+}(0h1p)  \xrightarrow[]{} \mathrm{Fl}^{}(0h2p)\label{eq:z114_FS-CC_sceme}.
\\[10pt]    
\end{aligned}
\end{eqnarray}
Within the FS-CC approach, the sector~$0h0p$ provides the total binding energy of the corresponding state, while for the other sectors the energies are determined relative to it.
Therefore, the $n$-th ionization potentials, $I_n$, and the electron affinity, $\varepsilon$, can be obtained by the proper combinations of the lowest eigenvalues of the effective Hamiltonian in all the sectors.
\par
The FS-CCSD approximation is used to evaluate the main contribution from the correlations to the IPs and EAs.
Explicitly optimized basis sets are employed in the calculations.
Their construction is discussed below.
The largest number of primitive functions used is~$37s36p24d18f6g4h3i2k1l$.
The occupied shells~$4f5s5p5d6s6p5f6d7s$ are explicitly treated at the FS-CCSD level.
Generally, more than~$60$ electrons are included in the correlation treatment for each element.
The energy cutoff for virtual states has been set to~$170$~a.u.
The contributions from the deep-lying core electrons and from the virtual states with the energies larger than~$170$~a.u.\ are investigated by performing the calculations with the reduced number of the correlated electrons and virtual states.
The MS determinants are constructed from the spinors~$6d7s$ in the Rg and Cn calculations and the~$6d7s7p8s$ ones in the Nh and Fl calculations.
As an additional check, the practical independence of the results on the number of the spinors, which span the MS, is verified.
\par
The contribution from the triple cluster amplitudes is defined as the difference between the FS-CCSDT and FS-CCSD results.
For all the elements, the triple-amplitude correction is evaluated considering the~$6d7s$ electrons as the active ones and excluding the virtual states with the energies larger than~$25$~a.u.
Overall, the FS-CCSDT equations are solved for about~$250$ spinors, depending on the considered element.
\par
As it is implemented in DIRAC, the DF equations are solved in the finite basis set of the primitive Gaussian functions~$\lbrace \gamma^L_i(\zeta_i) \rbrace$, where~$i$ enumerates basis functions of a given angular momentum~$L$, and~$\zeta_i$ stands for the Gaussian-exponent parameter.
Common basis sets~\cite{1993_MalliG_PhysRevA, 1998_DyallK_TheorChemAcc, 2001_FaegriK_TheorChemAcc}, which are widely used in calculations, may not be saturated enough to provide a proper description of the particular correlation problem at the desired accuracy level.
These basis sets are usually subjected to some customization.
A standard basis-set improvement strategy is to modify the basis set by including in it additional basis functions in the valence region in order to improve the accuracy of the electron-correlation description in this region.
\par
A different optimization procedure is used in the present work.
Previously, it was employed to calculate the EA of Og in Ref.~\cite{2021_KaygorodovM_PhysRevA}, where thorough details on the scheme can be found.
The procedure is based on the FS-CCSD calculations with the generalized relativistic effective core potential (GRECP)~\cite{1999_TitovA_IntJQuantumChem, 2020_MosyaginN_IntJQuantChem}, which significantly decreases the computational cost of the calculations by replacing core electrons with an effective pseudopotential.
In the present work, the basis set for each element under consideration is optimized with respect to all the IPs and EA which are to be calculated according to the schemes given by Eqs.~(\ref{eq:z111_FS-CC_sceme})\,--\,(\ref{eq:z114_FS-CC_sceme}).
\par
The optimization algorithm is organized as follows~\cite{2021_KaygorodovM_PhysRevA,Mosyagin:00}.
For a particular system, the corresponding most comprehensive Dyall's all-electron basis set~\cite{2011_DyallK_TheorChemAcc,2012_DyallK_TheorChemAcc}, with the basis functions for~$L>4$ being removed, is used as a starting point.
Beginning with~$L=0$, a basis function~$\gamma^L(\zeta)$ with an adjustable parameter~$\zeta$ is added to the initial basis set and the global extremum of the IP and EA values with respect to the parameter~$\zeta$ is searched.
The value of~$\zeta^*$ which simultaneously delivers the largest in magnitude extremum to all the considered IP and EA values is permanently added to the basis set.
For the given~$L$, the procedure continues iteratively until the studied quantities become stable (up to the desired accuracy) with respect to the addition of a new basis function to the basis set.
Then, the procedure for this value of~$L$ is stopped and optimization of the functions for~$L+1$ is started according to the same scheme.
\par
The range of the parameter~$\zeta$ related to the spatial region where the valence electrons are localized is empirically found to be $\zeta\in[0.01,10]$.
The basis functions $\gamma^L(\zeta)$ with the parameter $\zeta$ beyond this interval yield a negligible contribution to the investigated quantities.
\par
The effective core-potential integral-evaluation code employed in DIRAC allows one to use basis functions with the angular momenta up to~$L=6$ (the $i$-type functions).
To construct the optimized basis set with the maximum angular momentum~$L>6$, we use actually the same approach, but the GRECP Hamiltonian~$H_{\mathrm{GRECP}}$ is replaced with the Hamiltonian~$H_{\mathrm{X2Cmmf}}$ for which there is no such a restriction.
In addition, we have double checked the quality of the GRECP-optimized basis set by repeating the same optimization scheme but with the Hamiltonian~$H_{\mathrm{X2Cmmf}}$ employed instead of the~$H_{\mathrm{GRECP}}$ one.
This provides us a possibility to verify the established during the GRECP-optimization procedure uncertainty associated with the basis-set saturation error.
\par
Throughout all the CI-DFS calculations, the~$6d7s7p$ electrons are considered as the active ones, and the SD excitations are taken into account.
The convergence of the results with respect to the number of the virtual orbitals is studied.
\par
The CI-DFS method is used to calculate the Gaunt-interaction correction.
This contribution is evaluated as the difference between the values obtained with the Hamiltonians~$H_{\mathrm{DCG}}$ and~$H_{\mathrm{DC}}$.
In the same way, the correction associated with the retardation operator~$V^{\mathrm{R}}$ is calculated.
The QED corrections are also evaluated employing the CI-DFS method according to the scheme described in Sec.~\ref{sec:methods}, namely, as the difference of the energies obtained with and without the model-QED operator~$V^{\mathrm{QED}}$ included into the related calculations.
In this regard, we should mention the method presented in  Ref.~\cite{2021_SkripnikovL_JChemPhysA}, where the model-QED operator~$V^{\mathrm{QED}}$ was included into the FS-CC calculations.
In addition, the QED effects were incorporated into the atomic version of the FS-CC method in Ref.~\cite{2017_PastekaL_PhysRevLett}.

\section{Results and discussion}\label{sec:results}
We begin the discussion with the general results on the Gaunt and QED corrections.
Within the FS-CCSD method, the Gaunt-interaction correction is evaluated as the difference of the results obtained with and without the operator~$V^{\mathrm{G}}$ included into the Hamiltonian~$H_{\mathrm{X2Cmmf}}$.
Alternatively, it is calculated using the CI-DFS method.
The comparison of the Gaunt-interaction correction evaluated by the two different methods for the considered SHEs is presented in Table~\ref{tab:gaunt_comparison}.
\begin{table}[hbtp]
\centering
\caption{Comparison of the Gaunt-interaction corrections to the $n$-th ionization potentials, ${I_n}$, and electron affinities, ${\epsilon}$, of the SHEs with~$111\leq Z \leq 114$ evaluated with the FS-CCSD method using the exact two-component Hamiltonian~$H_{\mathrm{X2Cmmf}}$ and the CI-DFS method using the relativistic Dirac-Coulomb-Gaunt Hamiltonian (eV).}
\begin{tabular}{
c
c
S[table-format=-2.4]
S[table-format=-2.4]
}

\toprule
\multicolumn{1}{c}{$Z$} & \multicolumn{1}{c}{Quantity} & \multicolumn{1}{c}{FS-CCSD} & \multicolumn{1}{c}{CI-DFS}\\
\midrule
\multirow{2}{*}{111} & ${I_1\mathrm{(Rg)}}$      & 0.026 & 0.027 \\
                     & ${\epsilon\mathrm{(Rg)}}$ & 0.024 & 0.029 \\
\\[-8pt]
\multirow{2}{*}{112} & ${I_2\mathrm{(Cn)}}$      & 0.026 & 0.029 \\
                     & ${I_1\mathrm{(Cn)}}$      & 0.027 & 0.030 \\
\\[-8pt]
\multirow{4}{*}{113} & ${I_3\mathrm{(Nh)}}$      & -0.054 & -0.060 \\
                     & ${I_2\mathrm{(Nh)}}$      & -0.053 & -0.049 \\
                     & ${I_1\mathrm{(Nh)}}$      & -0.043 & -0.046 \\
                     & ${\epsilon\mathrm{(Nh)}}$ & -0.017 & -0.028 \\
\\[-8pt]
\multirow{4}{*}{114} & ${I_4\mathrm{(Fl)}}$      & -0.076 & -0.075 \\
                     & ${I_3\mathrm{(Fl)}}$      & -0.069 & -0.062 \\
                     & ${I_2\mathrm{(Fl)}}$      & -0.068 & -0.082 \\
                     & ${I_1\mathrm{(Fl)}}$      & -0.044 & -0.059 \\                     
\bottomrule
\end{tabular} 
\label{tab:gaunt_comparison}
\end{table}
\par
It can be seen from Table~\ref{tab:gaunt_comparison} that our FS-CCSD and CI-DFS results agree well with each other.
The maximum deviation between these results does not exceed~$0.015$~eV that is completely covered by the uncertainty of the FS-CCSDT values for the IPs and EAs.
The latter uncertainty is associated with the basis-set incompleteness and estimated to be several tenths of meV; see bellow the corresponding discussion for details.
The found agreement ensures us that the corrections obtained by means of the CI-DFS method can be added to the FS-CC results.
For the Gaunt-interaction corrections, the FS-CCSD results are considered by us to be more preferable than the CI-DFS ones, since a larger number of electrons are correlated in the FS-CCSD calculations.
\par
In the present work, the QED corrections are evaluated by using the CI-DFS method within the model-QED-operator approach.
However, for an illustrative purpose, we discuss also the evaluation of the QED correction at the one-electron level, namely, within the DF method.
The QED results obtained using the DF method are referred to as QED-DF.
To examine the effect of the model-QED-operator inclusion into the self-consistent DF equations on the studied quantities, we compare the expectation values of~$V^{\mathrm{QED}}$, obtained using the one-electron DF wave functions for the orbitals from which the ionization goes, with the QED corrections to the IPs and EAs which are obtained as the proper differences of the total DF energies calculated with and without the operator~$V^{\mathrm{QED}}$ included into the related DF equations.
The results of the first scheme can be interpreted as the \textit{direct} QED contribution to the considered quantities.
We call it the direct QED effect, since in this case the orbital relaxation which occurs due to the different electronic structure of the initial and final states is not taken into account.
On the other hand, the second scheme besides the direct QED effect accounts for an \textit{indirect} QED contribution from the orbital-relaxation effects which arise from the changes in the electronic structure as well as the orbital-relaxation effects due to including the model-QED potential into the DF equations.
The QED correction obtained by using the second scheme is referred to as the total QED-DF correction.
The corresponding results are presented in Table~\ref{tab:qed_comparison}.
\begin{table}[H]
\centering
\caption{Comparison of the direct and total QED contributions to the $n$-th ionization potentials, ${I_n}$, and electron affinities, ${\epsilon}$, of the SHEs with~$111\leq Z \leq 114$ (QED-DF) evaluated within the Dirac-Fock approximation using the model QED operator~$V^{\mathrm{QED}}$ (eV).
The direct QED-DF contribution is defined as the expectation value of~$V^{\mathrm{QED}}$ with the one-electron DF wave function for the orbital from which the ionization goes.
The total QED-DF contribution is defined as the proper difference of the total DF energies obtained with and without the operator~$V^{\mathrm{QED}}$ included into the DF self-consistent equations. See text for the details.
}

\begin{tabular}{
c
c
S[table-format=-1.4]
S[table-format=-1.4]
}

\toprule
\multicolumn{1}{c}{$Z$} & \multicolumn{1}{c}{Quantity} & \multicolumn{1}{c}{Direct QED-DF} & \multicolumn{1}{c}{Total QED-DF}\\
\midrule
\multirow{2}{*}{111} & ${I_1\mathrm{(Rg)}}$      & -0.0050 & 0.0230 \\
                     & ${\epsilon\mathrm{(Rg)}}$ & -0.0045 & 0.0216 \\
\\[-8pt]
\multirow{2}{*}{112} & ${I_2\mathrm{(Cn)}}$      & -0.0061 & 0.0259 \\
                     & ${I_1\mathrm{(Cn)}}$      & -0.0056 & 0.0244 \\
\\[-8pt]
\multirow{4}{*}{113} & ${I_3\mathrm{(Nh)}}$      & -0.0913 & -0.0926 \\
                     & ${I_2\mathrm{(Nh)}}$      & -0.0827 & -0.0779 \\
                     & ${I_1\mathrm{(Nh)}}$      & -0.0128 & -0.0017 \\
                     & ${\epsilon\mathrm{(Nh)}}$ & -0.0092 &  0.0019 \\
\\[-8pt]
\multirow{4}{*}{114} & ${I_4\mathrm{(Fl)}}$      & -0.1096 & -0.1097 \\
                     & ${I_3\mathrm{(Fl)}}$      & -0.1006 & -0.0948 \\
                     & ${I_2\mathrm{(Fl)}}$      & -0.0192 & -0.0057 \\
                     & ${I_1\mathrm{(Fl)}}$      & -0.0162 & -0.0021 \\                     
\bottomrule
\end{tabular} 
\label{tab:qed_comparison}
\end{table}
\par
From Table~\ref{tab:qed_comparison} it can be seen that in the cases of Rg and Cn the direct QED-DF contributions are several times in magnitude smaller than the total QED-DF ones, and have the opposite sign.
On the contrary, in the case of ${I_1\mathrm{(Nh)}}$, ${\epsilon\mathrm{(Nh)}}$, ${I_2\mathrm{(Fl)}}$, and  ${I_1\mathrm{(Fl)}}$ the direct QED-DF contributions turn out to be several times larger than the related total QED-DF ones.
These results indicate not the smallness or/and insignificance of the total QED-DF corrections but rather the fact that the proper treatment of the indirect QED-DF contributions associated with the relaxation effects (which are mostly due to the differences in the electronic structure of the charged state) may lead to a partial or in some cases to considerable cancellation of the direct QED-DF contributions.
Similar studies were performed in Ref.~\cite{2016_TupitsynI_PhysRevLett} for multivalent heavy ions and in Ref.~\cite{2020_ShabaevV_PhysRevA} for fluorine-like ions.
\par 
The QED-DF values are improved by using the CI-DFS method combined with the model-QED operator, see Sec.~\ref{sec:methods} for details.
For each SHE under consideration, the final QED corrections as well as the other contributions are presented in the forthcoming subsections.

\subsection{Roentgenium (Z=111, Rg)}\label{subsec:z111_results}
The results for the first IP and EA of roentgenium are presented in Table~\ref{tab:z111}.
We obtained that the ground-state configurations for the~Rg$^-$ anion is~$6d^{10} 7s^2$ (for brevity, here and below the common part~$\mathrm{[Rn]}5f^{14}$ of all
the configurations is omitted), for neutral Rg is $6d^{9}7s^2\,\,J=5/2$, and for Rg$^+$ ion is $6d^8 7s^2\,\,J=4$, in agreement with the previous FS-CCSD~\cite{1994_EliavE_PhysRevLett} and CI\texttt{+}PT~\cite{2016_DzubaV_PhysRevA,2020_LackenbyB_PhysRevA101} calculations.
The inclusion of the triple cluster amplitudes does not change the order of the levels, however, the corresponding correction to~$I_1(\mathrm{Rg})$ and~$\epsilon(\mathrm{Rg})$ turns out to be important. 
It amounts to~$-0.10(3)$~eV for the EA and~$-0.40(11)$~eV for the IP.
The uncertainty of the triple-amplitude correction was conservatively estimated by varying the virtual-states energy cutoff.
Both QED and Gaunt-interaction corrections for the EA and IP of Rg are about~$0.02$ eV, they are fully covered by the uncertainty of the FS-CC correlation treatment.
The retardation and the frequency-dependent Breit-interaction corrections are one and two orders of magnitude smaller, respectively, than the QED and Gaunt-interaction corrections.
\begin{table}[H]
\centering
\caption{Contributions to the first ionization potential,~$I_1(\mathrm{Rg})$, and electron affinity,~$\epsilon(\mathrm{Rg})$, of roentgenium~(eV).
The ground-state configuration of the neutral Rg atom is~$6d^{9}7s^2\,\,J=5/2$, the ground-state configurations of the ion/anion are given in the header.}
\begin{tabular}{
l
S[table-format=-3.5(1)]
S[table-format=-3.5(1)]
}

\toprule
\multirow{2}{*}{Contribution} & ${I_1(\mathrm{Rg})}$ & ${\epsilon(\mathrm{Rg})}$ \\
& ${6d^{8}7s^2\,\,J=4}$ & ${6d^{10}7s^2}$ \\ 
\midrule
FS-CCSD & 11.03(6) & 1.97(5) \\
Triples & -0.40(11)  & -0.10(3) \\
Gaunt & 0.026  & 0.024 \\
Retardation & -0.002  & -0.002 \\
Freq.-dep. Breit   & -0.0006 & 0.0001  \\
QED   & 0.021    & 0.019  \\
\midrule
Total & 10.67(13) & 1.91(6) \\
\midrule
Eliav \textit{et al.}~\cite{1994_EliavE_PhysRevLett} & 10.60 & 1.565 \\
Dzuba~\cite{2016_DzubaV_PhysRevA} & 12.2(11) & \\
Lackenby \textit{et al.}~\cite{2020_LackenbyB_PhysRevA101} & 11.175 & \\
\bottomrule
\end{tabular} 
\label{tab:z111}
\end{table}
\par
The main difference between the present FS-CC calculations and those reported in Ref.~\cite{1994_EliavE_PhysRevLett} consists in a more advanced basis set used here and in the inclusion of the triple cluster amplitudes in a nonperturbative, fully iterative manner~\cite{2020_OleynichenkoA_Symmetry}.
At the FS-CCSD level, our result for~$\epsilon(\mathrm{Rg)}$ differs from the value presented in Ref.~\cite{1994_EliavE_PhysRevLett} by about 25\%.
We managed to reproduce the values of Ref.~\cite{1994_EliavE_PhysRevLett} using the basis set employed there.
During the basis-construction procedure we found that both IP and EA of Rg increase in magnitude when the quality of the basis set is improved.
Since our values for~$I_1(\mathrm{Rg})$ and~$\epsilon(\mathrm{Rg})$ are larger by about~$0.4$~eV, we conclude that the difference between the results is apparently due to the lack of the indispensable basis functions used in that work.
Moreover, we note that our full Breit-interaction (Gaunt plus retardation) correction is in good agreement with the related results of Ref.~\cite{1994_EliavE_PhysRevLett}.
For~$I_1(\mathrm{Rg})$, the value of Ref.~\cite{1994_EliavE_PhysRevLett} is~$0.03$~eV and the present result is~$0.024$~eV, while for~$\epsilon(\mathrm{Rg})$ the value of Ref.~\cite{1994_EliavE_PhysRevLett} is~$0.023$~eV and our result is~$0.022$~eV.
The results obtained within the CI\texttt{+}PT calculations~\cite{2020_LackenbyB_PhysRevA101} are in reasonable agreement with our data, providing the triple-cluster-amplitude contribution is excluded from our total value.

\subsection{Copernicium (Z=112, Cn)}\label{subsec:z112_results}
For neutral and singly ionized copernicium the obtained ground-state configurations are $6d^{10} 7s^2$ and $6d^{9} 7s^2\,\,J=5/2$, respectively, in accordance with the previous FS-CCSD~\cite{1995_EliavE_PhysRevA} and CI\texttt{+}PT~\cite{2020_LackenbyB_PhysRevA101} calculations.
For $\mathrm{Cn}^{2+}$ ion, using the FS-CCSDT approach, we obtained that the~$J=4$ level of the configuration~$6d^{8} 7s^2$ has the minimal energy among the other levels, as it was previously reported in Refs.~\cite{1995_EliavE_PhysRevA,2020_LackenbyB_PhysRevA101}.
Thus, the ground-state configuration of Cn$^{2+}$ coincides with the ground-state configuration of Rg$^{+}$.
\begin{table}[H]
\centering
\caption{Contributions to the first two ionization potentials,~$I_n(\mathrm{Cn})$ with~$n=1,2$,  of copernicium~(eV).
The ground-state configuration of the neutral Cn atom is~$6d^{10}7s^2$, the ground-state configurations of the ions are given in the header.}
\begin{tabular}{
l
S[table-format=-2.6(1)]
S[table-format=-5.6(1)]
}

\toprule
\multirow{2}{*}{Contribution} & ${I_2(\mathrm{Cn})}$ & ${I_1(\mathrm{Cn})}$ \\
& ${6d^{8}7s^2\,\,J=4}$ & ${6d^{9}7s^2\,\,J=5/2}$ \\ 
\midrule
FS-CCSD & 22.54(6) & 12.00(5) \\
Triples & -0.18(9) & -0.03(2) \\
Gaunt & 0.026 & 0.027 \\
Retardation & -0.001 & -0.002 \\
Freq.-dep. Breit &-0.001 & -0.0006 \\
QED & 0.025 & 0.022 \\
\midrule
Total & 22.41(11) & 12.02(5) \\
\midrule
Eliav \textit{et al.}~\cite{1995_EliavE_PhysRevA} & 22.49 & 11.97 \\
Nash~\cite{2005_NashC_JPhysChemA} & & 11.675 \\
Yu \textit{et al.}~\cite{2007_YuY_EurPhysJD} & 21.98 & 11.73 \\
Hangele \textit{et al.}~\cite{2012_HangeleT_JChemPhys} & 21.989 & 11.353 \\
Dzuba~\cite{2016_DzubaV_PhysRevA} & & 13.1(11) \\
Lackenby \textit{et al.}~\cite{2020_LackenbyB_PhysRevA101} & 22.84 & 12.14 \\
\bottomrule
\end{tabular}
\label{tab:z112}
\end{table}
\par
In Table~\ref{tab:z112}, the contributions to the first, $I_1(\mathrm{Cn})$, and second, $I_2(\mathrm{Cn})$, IPs of copernicium are given.
For~$I_1(\mathrm{Cn})$, the triple-cluster-amplitude correction turns out to be about~$-0.03(2)$~eV, which is smaller than the FS-CCSD level of uncertainty.
On the contrary, the triple-amplitude correction is large for the ionization to the~$J=4$ level of Cn$^{2+}$ and amounts to about~$-0.18(9)$~eV.
Nevertheless, it does not change the order of the levels.
The full Breit interaction and QED corrections contribute about~$0.02$\,--\,$0.03$~eV to the IPs of copernicium, and they are roughly the same as for the case of Rg.
\par
Omitting the triple cluster amplitudes and QED corrections, our results for both first and second IPs are in good agreement with the previous FS-CCSD calculations~\cite{1995_EliavE_PhysRevA}.
However, for the first IP our full Breit-interaction contribution, which is~$0.025$~eV, disagrees with the one from Ref.~\cite{1995_EliavE_PhysRevA}, which is~$-0.022$~eV, \textit{i.e.} it has almost the same magnitude but the opposite sign.
On the other hand, our value for the Gaunt-interaction correction is in reasonable agreement with the result of Ref.~\cite{2010_ThierfelderC_PhysRevA}, which is~$0.031$~eV.
The obtained QED correction, $0.022$~eV, agrees well with the value from Ref.~\cite{2010_ThierfelderC_PhysRevA}, which is~$0.023$ eV.
\par
Our total results excluding the triple-cluster-amplitude corrections for $I_1(\mathrm{Cn})$ and $I_2(\mathrm{Cn})$ are 1\% and 2\% smaller than the ones from Ref.~\cite{2020_LackenbyB_PhysRevA101}, respectively.
The addition of the triple-amplitude corrections, which are negative for all cases, only increases the difference.
Finally, our results are in agreement with the results of Ref.~\cite{2007_YuY_EurPhysJD} within the $0.5$~eV uncertainty estimated by the authors of Ref.~\cite{2007_YuY_EurPhysJD}.
The corresponding values were obtained using the GRASP92~\cite{1996_ParpiaF_CPC} implementation of the MCDF method.

\subsection{Nihonium (Z=113, Nh)}\label{subsec:z113_results}
The different charge states of nihonium have the following ground-state configurations: $6d^{10}7s^27p^1$ for the neutral Nh, $6d^{10}7s^27p^2$ for the Nh$^-$ anion, and
$6d^{10}7s^2$, $6d^{10}7s^1$, and $6d^{10}$ for singly, doubly, and triply ionized Nh, respectively.
The occupation of the $6d$ shell remains unchanged for the first three ionization processes in Nh, whereas in cases of Rg and Cn the electrons are ionized namely from the $6d$ shell.
The obtained ground-state configurations are in agreement with the previous results of Refs.~\cite{1996_EliavE_PhysRevA, 2016_DzubaV_HyperfInteract237}.
The inclusion of the nonperturbative triple cluster amplitudes does not change the ground-state configurations as well.
\par
The contribution of the triple amplitudes to the IPs and EA of Nh is about a few tenths of meV, being positive for~$\epsilon(\mathrm{Nh})$ and~$I_1(\mathrm{Nh})$ and negative for~$I_2(\mathrm{Nh})$ and~$I_3(\mathrm{Nh})$.
The QED correction to~$\epsilon(\mathrm{Nh})$ and~$I_1(\mathrm{Nh})$ is negligible in comparison with the uncertainty of the FS-CC correlation treatment since this correction is mainly determined by the~$7p$ electrons.
However, this is not the case for~$I_2(\mathrm{Nh})$ and~$I_3(\mathrm{Nh})$, where the~$7s$ electrons are ionized.
For the second and third IPs, the QED corrections amount to about~$-0.08$~eV and~$-0.09$~eV, respectively.
\begin{table}[H]
\centering
\caption{Contributions to the first three ionization potentials, $I_n(\mathrm{Nh})$ with~$n=1$\,--\,$3$, and electron affinity, $\epsilon(\mathrm{Nh})$, of nihonium (eV).
The ground-state configuration of the neutral Nh atom is~$6d^{10}7s^27p^1$, the ground-state configurations of the ions/anion are given in the header.}

\begin{tabular}{
l
S[table-format=-1.2(1)]
S[table-format=-1.2(1)]
S[table-format=-1.2(1)]
S[table-format=-1.2(2)]
}

\toprule
\multirow{2}{*}{Contribution} & 
${I_3(\mathrm{Nh})}$ & ${I_2(\mathrm{Nh})}$ & ${I_1(\mathrm{Nh})}$ & ${\epsilon(\mathrm{Nh})}$ \\
& ${6d^{10}}$ & ${6d^{10}7s^1}$ & ${6d^{10}7s^2}$ & ${6d^{10}7s^27p^2}$\\ 
\midrule
FS-CCSD & 33.52(6) & 24.00(5) & 7.49(4) & 0.71(3) \\
Triples & -0.01(3) & -0.03(3) & 0.04(2) & 0.03(2) \\
Gaunt & -0.054 & -0.053 & -0.043 & -0.017 \\
Retardation & 0.004 & 0.003 & 0.004 & 0.003 \\
Freq.-dep. Breit & -0.005 & -0.005 & -0.003 & -0.002 \\
QED & -0.088 & -0.078 & -0.001 & 0.002 \\
\midrule
Total & 33.37(7) & 23.84(6) & 7.49(5) & 0.73(4) \\
\midrule
Eliav \textit{et al.}~\cite{1996_EliavE_PhysRevA} & 33.47 & 23.96 & 7.306 & 0.68(5) \\
Pershina \textit{et al.}~\cite{2008_PershinaV_JPhysChemA} & & & 7.420 & \\
Hangele \textit{et al.}~\cite{2012_HangeleT_JChemPhys} & &23.627 & 7.278 & \\
\makecell[cl]{Demidov and \\ Zaitsevkii}~\cite{2015_DemidovY_CPL} & & & 7.44 & \\
\makecell[cl]{Dzuba and \\ Flambaum}~\cite{2016_DzubaV_HyperfInteract237} & 33.5 & 23.6 & 7.37 &  \\
Guo \textit{et al.}~\cite{2022_GuoY_ArXiv220201294} &  &  & 7.569(48) & 0.776(30) \\
\bottomrule
\end{tabular}
\label{tab:z113}
\end{table}
\par
During the basis-set quality-improvement procedure we found that the values~$\epsilon(\mathrm{Nh})$ and~$I_1(\mathrm{Nh})$ increase in magnitude.
This could be an explanation why our FS-CCSD value for $I_1(\mathrm{Nh})$ combined with the full Breit-interaction correction is about $0.15$~eV larger than the corresponding value from Ref.~\cite{1996_EliavE_PhysRevA}.
The total Breit-interaction correction is in good agreement with the results of Ref.~\cite{1996_EliavE_PhysRevA} for all the considered quantities.
For~$I_1(\mathrm{Nh})$, our value,~$-0.039$~eV, for the full Breit-interaction correction also agrees with Ref.~\cite{2010_ThierfelderC_PhysRevA}, where~$-0.046$~eV for this correction was obtained by including the operator~$V^{\mathrm{G}}+V^{\mathrm{R}}$ into the self-consistent procedure.
\par
The total values for the EA and IPs of Nh are in agreement with the CI+PT results of Ref.~\cite{2016_DzubaV_HyperfInteract237}, where the $1\%$ uncertainty for the energies was reported.
The result for~$I_1(\mathrm{Nh})$ agrees within the estimated uncertainty with the one obtained by Demidov and Zaitsevskii~\cite{2015_DemidovY_CPL} who employed the scalar relativistic CC-SD(T) method, with the spin-dependent relativistic corrections being evaluated using the two-component relativistic density functional theory.
Our results for~$I_1(\mathrm{Nh})$ and~$I_2(\mathrm{Nh})$, as a whole, agree with the results of Ref.~\cite{2012_HangeleT_JChemPhys}.
The present~$I_1(\mathrm{Nh})$ and~$\varepsilon(\mathrm{Nh})$ values agree also with the very recent results from Ref.~\cite{2022_GuoY_ArXiv220201294} which were obtained by the use of the CC method including the QED corrections.

\subsection{Flerovium (Z=114, Fl)}\label{subsec:z114_results}
The obtained ground-state configuration for the Fl atom is~$6d^{10} 7s^2 7p^2$.
The order of the electron detachment in the Fl element is the following: at first, two~$7p_{1/2}$ electrons are ionized, then two~$7s$ ones, leaving the ion Fl$^{4+}$ with the~$6d^{10}$ ground-state configuration.
The obtained ground-state configurations for the Fl atom and its ions are in agreement with the previous findings made in Refs.~\cite{2001_LandauA_JChemPhys, 2008_YuY_JChemPhys,2016_DzubaV_HyperfInteract237}: the triple-amplitude contribution does not change the ground-state configurations for the Fl ions.
The results for the IPs of flerovium are presented in Table~\ref{tab:z114}.
\begin{table}[H]
\centering
\caption{Contributions to the first four ionization potentials,~$I_n(\mathrm{Fl})$ with~$n=1$\,--\,$4$, of flerovium~(eV).
The ground-state configuration of the neutral Fl atom is~$6d^{10}7s^27p^2$, the ground-state configurations of the ions are given in the header.}
\begin{tabular}{
l
S[table-format=-1.2(1)]
S[table-format=-1.2(1)]
S[table-format=-1.2(1)]
S[table-format=-1.2(1)]
}

\toprule
\multirow{2}{*}{Contribution} & 
${I_4(\mathrm{Fl})}$ & ${I_3(\mathrm{Fl})}$ & ${I_2(\mathrm{Fl})}$ & ${I_1(\mathrm{Fl})}$ \\
& ${6d^{10}}$ & ${6d^{10}7s^1}$ & ${6d^{10}7s^2}$ & ${6d^{10}7s^27p^1}$\\ 
\midrule
FS-CCSD & 46.24(6) & 35.78(6) & 16.95(5) & 8.69(5) \\
Triples & -0.01(3) & -0.05(3) & 0.04(3) & 0.01(2) \\
Gaunt& -0.076 & -0.069 & -0.068 & -0.044 \\
Retardation & 0.006& 0.004 & 0.006 & 0.004 \\
Freq.-dep. Breit & -0.006 & -0.006 & -0.004 & -0.003 \\
QED & -0.105 & -0.093 & -0.004 & -0.002 \\
\midrule
Total & 46.05(7) & 35.56(7) & 16.92(6) & 8.65(5) \\
\midrule
Seth \textit{et al.}~\cite{1998_SethM_AngewChemIntEd} & 	& 35.52 & 16.55 & 8.36 \\
Landau \textit{et al.}~\cite{2001_LandauA_JChemPhys} & 46.272	& 35.739 & 16.871 & 8.539 \\
Nash~\cite{2005_NashC_JPhysChemA} & & & & 8.529 \\
Yu \textit{et al.}~\cite{2008_YuY_JChemPhys} & 46.57 & 35.82 & 17.22 & 8.28 \\
Hangele \textit{et al.}~\cite{2012_HangeleT_JChemPhys} & & 35.383 & 16.111 & 7.260 \\
\makecell[cl]{Dzuba and \\ Flambaum}~\cite{2016_DzubaV_HyperfInteract237} &  & & 17.00 & 8.37 \\
\bottomrule
\end{tabular}
\label{tab:z114}
\end{table}
\par
The triple-amplitude correction to~$I_2(\mathrm{Fl})$ and~$I_3(\mathrm{Fl})$ turns out to be tenths of meV, and these contributions are comparable with the FS-CCSD uncertainties.
It is interesting to note that for Fl this correction appears to be roughly the same as for Nh.
The QED corrections for~$I_3(\mathrm{Fl})$ and~$I_4(\mathrm{Fl})$, where the~$7s$ electrons are ionized, are found to be about~$-0.1$~eV.
On the contrary, for~$I_1(\mathrm{Fl})$ and~$I_2(\mathrm{Fl})$ the QED corrections are found to be negligible compared to the uncertainty due to the approximate treatment of electronic correlations.
The retardation and the frequency-dependent Breit corrections almost cancel each other yielding several-meV contributions to the IPs.
They are an order of magnitude smaller than the numerical uncertainty associated with the correlation treatment.
\par
Our FS-CCSD values combined with the Gaunt and retardation corrections agree well with the results of Ref.~\cite{2001_LandauA_JChemPhys}, which were obtained using the FS-CCSD method with the DCB Hamiltonian.
The~$I_1(\mathrm{Fl})$ value obtained in Ref.~\cite{2005_NashC_JPhysChemA} by means of the CCSD(T) method is also in agreement with our results.
Since the triple-cluster-amplitude contribution turns out to be small for~$I_1(\mathrm{Fl})$, the single-reference CCSD(T) method, where the triple amplitudes are evaluated pertubatively, seems to yield a reliable result in this case.
However, as in case of~Cn, the basis set used in Ref.~\cite{2005_NashC_JPhysChemA} was considerably smaller than the one employed in the present work.
\par
Our total results for the first and second potentials agree up to several percent with the values from Ref.~\cite{2016_DzubaV_HyperfInteract237}, which were obtained by means of the CI approach combined with the linearized CCSD method~\cite{2014_DzubaV_PhysRevA} and also include the Breit and QED corrections.
Furthermore, our total results turn out to be systematically larger by~$3-4\%$ than those from Ref.~\cite{2008_YuY_JChemPhys}, where the uncertainty was estimated to be~$2000$~cm$^{-1}$, which is about~$0.25$~eV.
With this uncertainty, our results agree better for~$I_2(\mathrm{Fl})$ and~$I_3(\mathrm{Fl})$ and worse for~$I_1(\mathrm{Fl})$ and $I_4(\mathrm{Fl})$.
The discrepancy may be a result of the fact that a configuration space employed in Ref.~\cite{2008_YuY_JChemPhys} was not saturated sufficiently.
Finally, we note that the results from Ref.~\cite{2012_HangeleT_JChemPhys} for Fl obtained using a pseudopotential approach deviate from the corresponding our results: the deviation is larger than~$1$~eV for~$I_1(\mathrm{Fl})$, about~$0.8$~eV for~$I_2(\mathrm{Fl})$, and about~$0.2$~eV for~$I_3(\mathrm{Fl})$.
The reason of this discrepancy is unclear to us.

\section{Conclusion}\label{sec:conclusion}
The ionization potentials and electron affinities of the SHEs with~$111 \leq Z\leq 114$ are calculated by means of the relativistic FS-CC method within the Dirac-Coulomb Hamiltonian.
To this end, the optimized basis sets of the primitive Gaussian exponents were constructed for each SHE considered.
For these systems for the first time, the nonperturbative triple-amplitude corrections are evaluated based on the implementation of Refs.~\cite{Oleynichenko_EXPT,EXPT_website,2020_OleynichenkoA_Symmetry}.
Especially in those cases, where the occupation of the~$d$ shell is changed during the ionization process, the previously unaccounted corrections due to the triple amplitudes turn out to be significant and even larger than the uncertainty of the FS-CCSD correlation treatment.
The Breit-interaction and frequency-dependent Breit-interaction corrections are calculated by means of the CI-DFS method~\cite{2003_TupitsynI_OptSpectrosc,2003_TupitsynI_PhysRevA,2005_TupitsynI_PhysRevA,2018_TupitsynI_PhysRevA}.
The QED corrections are also evaluated within the CI-DFS approach employing the model-QED operator~\cite{2013_ShabaevV_PhysRevA,2015_ShabaevV_CompPhysComm,*2018_ShabaevV_CompPhysComm}.
The corresponding contributions amount to about~$-0.1$~eV in the cases, where the ionization goes from the~$s$ shell.
\par
In most cases, the numerical uncertainty of our calculations does not exceed~$0.1$~eV.
It is mainly determined by the uncertainties associated with the finite size of the basis sets employed and the uncertainty arising from the core electrons uncorrelated within the FS-CCSDT model.
With the QED and triple-amplitude contributions being excluded, our results are in agreement with the previous FS-CCSD ones, which were obtained by Eliav and coworkers in a series of works~\cite{1994_EliavE_PhysRevLett,1995_EliavE_PhysRevA,1996_EliavE_PhysRevA,2001_LandauA_JChemPhys} where these contributions were not considered.
However, for roentgenium using the more advanced basis set we managed to significantly improve the electron-correlation description compared to the previous FS-CCSD calculations~\cite{1994_EliavE_PhysRevLett}.
Our predictions are also in reasonable agreement with the results from Refs.~\cite{2016_DzubaV_HyperfInteract237,2020_LackenbyB_PhysRevA101} obtained using the CI\texttt{+}PT method and taking into account the QED corrections.
However, for a reason that is unclear to us, the discrepancy of more than~$0.25$~eV takes place for the first ionization potential of Rg, the second ionization potential of Cn, and the first ionization potential of Fl.

\section{Acknowledgements}
The work is supported by the Ministry of Science and Higher Education of the Russian Federation within the~Grant~No.~075-10-2020-117.

\bibliographystyle{apsrev4-2}
\bibliography{main}
\end{document}